\documentstyle[12pt]{article}
\begin{document}
\title{Complex Patterns in a Simple System}
\author{John E. Pearson \\
Center for Nonlinear Studies \\ Los Alamos National Laboratory}
\maketitle
\begin{abstract}
Numerical simulations of a  simple reaction--diffusion
model reveal a surprising variety of irregular spatio--temporal patterns.
These patterns arise in response to finite--amplitude
perturbations. Some of them
resemble the steady irregular patterns discussed by Lee et al.
in this issue of {\it Science}~\cite{Lee}. Others consist of
spots which grow until they reach a critical
size at which time they divide in two. If, in some region,
the spots become over--crowded, all the spots in that region decay into
the uniform background.
\end{abstract}
\pagebreak
Patterns occur in nature at scales ranging from the developing Drosophila
embryo to the large--scale structure of the universe. At the familiar
mundane scales we see snowflakes, cloud streets, and sand ripples. We
see convective roll patterns in hydrodynamic experiments. We see regular
and almost regular patterns in the concentrations of chemically reacting
and diffusing systems.
As a consequence of the enormous
range of scales over which pattern formation occurs, the discovery
of any new pattern formation phenomenon is potentially of great scientific
interest. In this article, I describe patterns recently
observed in numerical experiments on a simple reaction--diffusion
model. These patterns are unlike any that have been
previously observed in theoretical or numerical studies.

The system is a variant of the autocatalytic Selkov model of
glycolysis~\cite{Sel}
and is due to Gray and Scott~\cite{GnS}.
A variety of spatio--temporal
patterns form in response to finite--amplitude
perturbations. The response of this model to such
perturbations was previously studied in one space dimension by Vastano et al.
who showed that steady spatial patterns could form even when the
diffusion coefficients were equal~\cite{V1}.
The response of the system in one space dimension is nontrivial
and depends both on the control parameters and on the
initial perturbation. It will be shown that
the patterns that occur in two dimensions range
from the well--known regular hexagons to irregular steady patterns
similar to those recently observed by Lee et al.~\cite{Lee},
to chaotic spatio--temporal patterns. For the ratio of diffusion
coefficients used, there are no stable Turing patterns.

Most work in this field in the past has focused
on pattern formation from a spatially uniform state that is near the
transition from linear stability to linear instability. With this restriction,
standard bifurcation--theoretic tools such as amplitude equations have been
developed and used  with considerable success~\cite{Seg}--\cite{New}.
It is unclear whether the
patterns presented in this article will yield to these now--standard
technologies.

The Gray--Scott model corresponds to the
following two reactions:
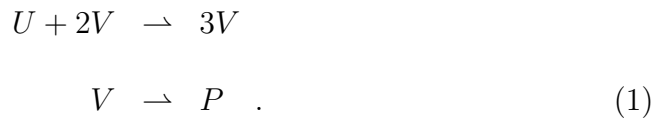
\begin{eqnarray}
           U + 2V &\rightharpoonup & 3V \cr
             \label{eq:React} \cr
             V    &\rightharpoonup &  P\quad .
\end{eqnarray}
Both reactions are irreversible so $P$ is an inert product.
A nonequilibrium constraint is represented by a feed
term for $U$. Both $U$ and $V$ are removed by the feed process.
The resulting reaction--diffusion equations in dimensionless
units are
\begin{eqnarray}
{\partial U \over \partial t} &=& D_u \nabla^2 U -UV^2 + F(1-U) \cr
       \label{eq:RD}   \cr
{\partial V \over \partial t} &=& D_v \nabla^2 V +UV^2 - (F+k)V  \quad ,
\end{eqnarray}
where $k$ is the dimensionless rate constant of the second reaction
and $F$ is the dimensionless feed rate.
The system size is $2.5\times 2.5$,
$D_u = 2\times 10^{-5}$ and $D_v = 10^{-5}$. The boundary conditions are
periodic. Before the numerical results are presented, let us consider the
the behavior of the reaction kinetics which are described by the ordinary
differential equations that result upon dropping the diffusion terms in
Eq.~\ref{eq:RD}.

Figure 1 is a phase diagram. A trivial steady state $U=1, V=0$ exists and
is linearly stable for all $F$ and $k$. In the region bounded above by
the solid line and below by the dotted line, the system is bistable.
For fixed $k$ the nontrivial stable uniform solution loses stability
via saddle--node bifurcation as $F$ is increased through the upper
solid line or by Hopf bifurcation as $F$ is decreased through the
dotted line~\cite{FOOT}.  In the case at hand the bifurcating periodic
solution is stable for $k<.035$ and unstable for $k>.035$. There is no
periodic orbit for parameter values outside the region enclosed by the
solid line.

The simulations are forward
Euler integration of the finite difference equations resulting from
discretization of the diffusion operator. The spatial mesh consists of
$256 \times 256$ grid points. The time step used is 1.
Spot checks made with meshes as large as $1024 \times 1024$
and time steps as small as $.01$ produce no qualitative difference in the
results.

Initially, the entire system was placed in
the trivial state $(U=1,V=0)$. Then the $20 \times 20$ mesh point area located
symmetrically about the center of
the grid was perturbed to $(U=1/2, V=1/4)$.  These conditions were then
perturbed with $\pm 1\%$ random noise in order to break the square
symmetry. The system was then integrated for $200,000$
time steps and an image saved.
In all cases, the initial disturbance propagated
outward from the central square until the entire grid was affected by the
initial square perturbation. The propagation was wavelike, with the leading
edge of the perturbation moving with an approximately constant velocity.
Depending on the parameter values, it took on the order of $10,000$--$20,000$
time steps for the initial perturbation to spread over the entire grid.
The propagation velocity of the initial perturbation is thus on the order
of $1\times 10^{-4}$ space units per time unit. After the initial period
during which the perturbation spreads, either the system goes into a time
dependent state, or to an essentially steady state.

Figures 2 and 3 are phase diagrams. One can view Figure 3 as a map and
Figure 2 as the key to the map.  The twelve patterns illustrated
in Figure 2 and are designated by Greek letters
and will be referred to
as ``pattern $\alpha$'', ``pattern $\beta$'', etc.
The color indicates the concentration of $U$ with red
representing $U=1$ and blue representing $U\approx .2$.
Yellow is intermediate to red and blue.
In Figure 3, the Greek characters indicate the type of pattern
found at that point in parameter space.
There are two additional
symbols in Figure 3, ``R'''s
and ``B'''s indicating spatially uniform red and blue states respectively.
The red state corresponds to $(U=1,V=0)$ and the blue state depends on
the exact parameter values but corresponds roughly to $(U=.3, V=.25)$.

Pattern $\alpha$ is time--dependent and consists of fledgling spirals which
are constantly colliding and annihilating each other: full spirals never
form. Pattern $\beta$ is time--dependent and consists of what
is generally called phase turbulence~\cite{KURA} which occurs in the vicinity
of a Hopf bifurcation to a stable periodic orbit. The medium is unable to
synchronize so the phase of the oscillators varies as a function of position.
In the present case the small amplitude periodic orbit that bifurcates is
unstable. Pattern $\gamma$ is time dependent. It consists primarily of stripes
as illustrated but there are small localized regions that oscillate with a
relatively high frequency $\approx 10^{-3}$. The active regions disappear but
new ones always appear elsewhere. In Figure 2 there is an active region near
the top center of pattern $\gamma$. Pattern $\delta$ consists of
regular hexagons except for apparently stable defects.

Pattern $\eta$ is time dependent: a few of the stripes
oscillate without apparent decay, but the remainder of the pattern remains
time independent.

Pattern $\iota$ is time independent and was observed for only a single
parameter value.

Patterns $\theta$, $\kappa$, and $\mu$ resemble those observed by
Lee et al. and described in this issue of {\it Science}~\cite{Lee}.
When blue waves collided, they
stopped, as did the waves observed by Lee. In pattern $\mu$,
long stripes grew in length.
The growth was parallel to the stripes and took
place at the tips. If two distinct stripes were both growing
and were pointed directly at each other, it was always observed
that when the growing tips reached some critical separation distance, they
would alter their course so as not to collide. In patterns $\theta$ and
$\kappa$ the perturbations grew radially outward with a velocity that
was normal to the stripes. In this case if two stripes collided, they
simply stopped, as did those observed by Lee.
In simulations in one space
dimension, I have also observed fronts propagating towards each other that stop
when they reach a critical separation. This is fundamentally new behavior
for nonlinear waves.

Patterns $\varepsilon$, $\zeta$, and $\lambda$ are the most interesting
and share similarities. They consist of blue spots on a red or yellow
background. Pattern $\lambda$ is time independent and patterns $\varepsilon$
and $\zeta$ are time dependent. Note that spots occur only in regions
of parameter space where the
sole uniform steady state is the red state $(U=1, V=0)$. Thus, the
blue spots cannot exist for extended time unless there is a gradient
present. Since the gradient is required for the existence of the
spots, they must have a maximum size otherwise there would be blue
regions that were essentially gradient free. Such regions would
necessarily decay to the red state.
Note that these gradients are self--sustaining and {\it not}
imposed externally. After the initial perturbation,
the spots increase in number until they fill the system.
How do they increase their number?
After a spot is formed, it grows. When it achieves the critical
size the gradient is no longer sufficient to maintain the center of the
spot in the blue state so the center decays to red leaving two blue spots.
This process is illustrated in Figure 4. The subsequent evolution depends
on the control parameters. Pattern $\lambda$ remains in a steady state.
Pattern $\zeta$ remains time dependent but with long--range spatial order
except for local regions of activity. The active regions are not stationary.
At anyone instant they do not appear significantly different from pattern
$\zeta$ as illustrated in Figure 2 but the location of the red disturbances
changes with time.
Pattern $\varepsilon$ appears to have no long--range order either in time or
space. Once the system is filled with
blue spots, they can die due to over--crowding. This occurs when many spots
are crowded together and the gradient becomes insufficient to support them.
The spots in this extended region will collapse nearly simultaneously
leaving an irregular red hole. There are always more spots on the
boundary of any hole and after a few thousand time steps no sign of the
hole will remain. The spots on its border will have filled it.
Figures 4 and 5 illustrate the birth and death of the spots in
pattern $\varepsilon$.

Figure $6$
provides an estimate of the largest Liapunov exponent for
pattern $\varepsilon$. The system
(with $F=.02, k=.059$) was evolved for $20,000$ time steps. Then
a copy of the system was made and randomly perturbed
with noise on the order of $1\times 10^{-12}$. The perturbed state served
as the initial condition for the copy which was then evolved simultaneously
with the original. This was done $20$ times and an average was taken.
Figure 6 is a $Log_{10}$--linear plot of $D(t)$ versus
$t$ where
$$
D(t) = <|d|>
$$
and $|d|$ is the space average of the absolute value of the difference between
the perturbed solution and the unperturbed solution. The angular brackets
indicate that an ensemble average was taken. One can clearly see the
straight line positive slope
dependence indicating exponential growth of errors---the hallmark of chaos.
The Liapunov exponent is on the order of $1.5\times 10^{-3}$.
The Liapunov time (the inverse of the Liapunov exponent) is $660$ time steps.
It is roughly equal to the time it takes for a spot to reproduce as shown in
Figure 4. This time is also on the order of the time it takes for a molecule
to diffuse across one of the blue spots.

All of the patterns presented here arose in response to finite--amplitude
perturbations. The ratio of diffusion coefficients used was $2$. It is
now well known that Turing instabilities which lead to spontaneous pattern
formation cannot occur in systems in which all diffusion coefficients
are equal. For a comprehensive discussion of these issues see
Pearson~\cite{P1}-\cite{P2}.
For a discussion of Turing instabilities in the model at hand, see
Vastano et al~\cite{V2}. The only Turing patterns
that can occur bifurcate off the
nontrivial steady uniform state (the blue state).
With the ratio
of diffusion coefficients used here, they occur only in a narrow
parameter region in the vicinity of $(F=k=.0625)$ where
the line of saddle--node bifurcations coalesces with the line of Hopf
bifurcations. In the vicinity of this point it was shown that
the branch of small amplitude Turing patterns is
unstable~\cite{V2}. The smallest ratio of diffusion coefficients
that gives stable Turing patterns is about $D_u/D_v=2.8$.

No pattern formation occurred when the diffusion coefficients were equal
although target patterns were observed. The target patterns consisted
of an oscillating core located at the center of
the initially perturbed region. The core oscillated, apparently without decay,
continually launching waves into the medium. With equal diffusion coefficients,
no patterns formed in which the small asymmetries in the initial
conditions were amplified by the dynamics. This observation can probably
be understood in terms of the following fact. Nonlinear plane waves in two
dimensions cannot be destabilized by diffusion in the case that all diffusion
coefficients are equal~\cite{P4}.
During the initial stages of the evolution, the corners of the square
perturbation are rounded off. Then the perturbation
evolves as a radial wave, either inward or outward depending on the parameter
values. Such a wave cannot undergo spontaneous symmetry breaking unless the
diffusion coefficients are unequal. However, as demonstrated in the current
article, I found symmetry breaking over a wide range of parameter values
for a ratio of diffusion coefficients of $2$. Such a ratio is physically
reasonable even for small molecules in aqueous solution.

The similarity of the Gray--Scott model to the Selkov model of glycolysis
may be of biological relevance. Recently Hasslacher et al., have demonstrated
the plausibilty of sub--cellular chemical patterns via lattice--gas simulations
of the Selkov model~\cite{HAS}. The patterns discussed in the present
article are likely to be found in the Selkov model as well, and thus
it is plausible that they are relevant to glycolysis.

\pagebreak
\centerline{Figure Captions}
\bigskip
\begin{itemize}
\item{Figure 1:}
Phase diagram of the reaction kinetics: Outside the region bounded by the
solid line, there is a single spatially uniform state (called the
trivial state) $(U=1, V=0)$ that
is stable for all $(F, k)$.  Inside the region bounded by the
solid line, there are three spatially uniform steady states. Above the
dotted line and below the solid line, the system is bistable:
there are two linearly stable steady states in this region. As $F$
is decreased through the dotted line the nontrivial stable steady state
loses stability via Hopf bifurcation. The bifurcating periodic orbit is
stable for $k<.035$ and unstable for $k>.035$. No periodic orbit exists
for parameter values outside the region bounded by the solid line.
\medskip
\item{Figure 2:}
The key to the map. The patterns
shown in the figure are designated by Greek letters which are used in
Figure 3 to indicate the type of pattern found at a given point in
parameter space.
\medskip
\item{Figure 3:}
The map.
The Greek letters indicate that
a pattern similar to the pattern with the same Greek letter in Figure 2 was
found at
that point in parameter space. The ``B'''s indicate that the system evolved to
a uniform blue state. The ``R'''s indicate that the system evolved to
a uniform red state.
\medskip
\item{Figure 4:}
Time evolution of spot multiplication. This figure was produced in a
$256 \times 256$ simulation with physical dimensions of $.5 \times .5$
and a time step of $.01$. The times at which the figures were taken
is as follows: (A) $T=0$, (B) $T=350$, (C) $T=510$, (D) $T=650$.
\medskip
\item{Figure 5:}
Time evolution of pattern $\varepsilon$. The images are $250$ time
units apart. In the corners (which map to the same point in physical
space) one can see a yellow region in Figures A-C. It has decayed to
red in Figure D. In Figures A-B the center of the left border
has a red region which is nearly filled in Figure D.
\medskip
\item{Figure 6:}
$Log_{10}$--linear plot of $D$ vs $t$. The center line is the average
and the upper and lower lines are the average plus and minus the standard
deviation.
The straight line has a
slope corresponding to a positive Liapunov exponent of about
$1.5\times 10^{-3}$.
\end{itemize}


\begin{thebibliography}{Bordeaux}
\bibitem{FOOT}In this foot--note I describe briefly what is meant by the terms
saddle--node and Hopf bifurcations. Consider the ordinary differential
equation $d{\vec U}/dt = {\vec F}({\vec U}; \mu)$ where $\mu$ is a control
parameter.
A saddle--node bifurcation occurs when the number of distinct
real zeroes of ${\vec F}$ changes as $\mu$ is
varied. A Hopf bifurcation occurs when the system becomes linearly unstable
to oscillatory modes. When a system undergoes a saddle--node bifurcation
it settles at a new state which is far from the old state. At a Hopf
bifurcation, a branch of periodic solutions appears. This branch may be
either stable or unstable.
\end{thebibliography}
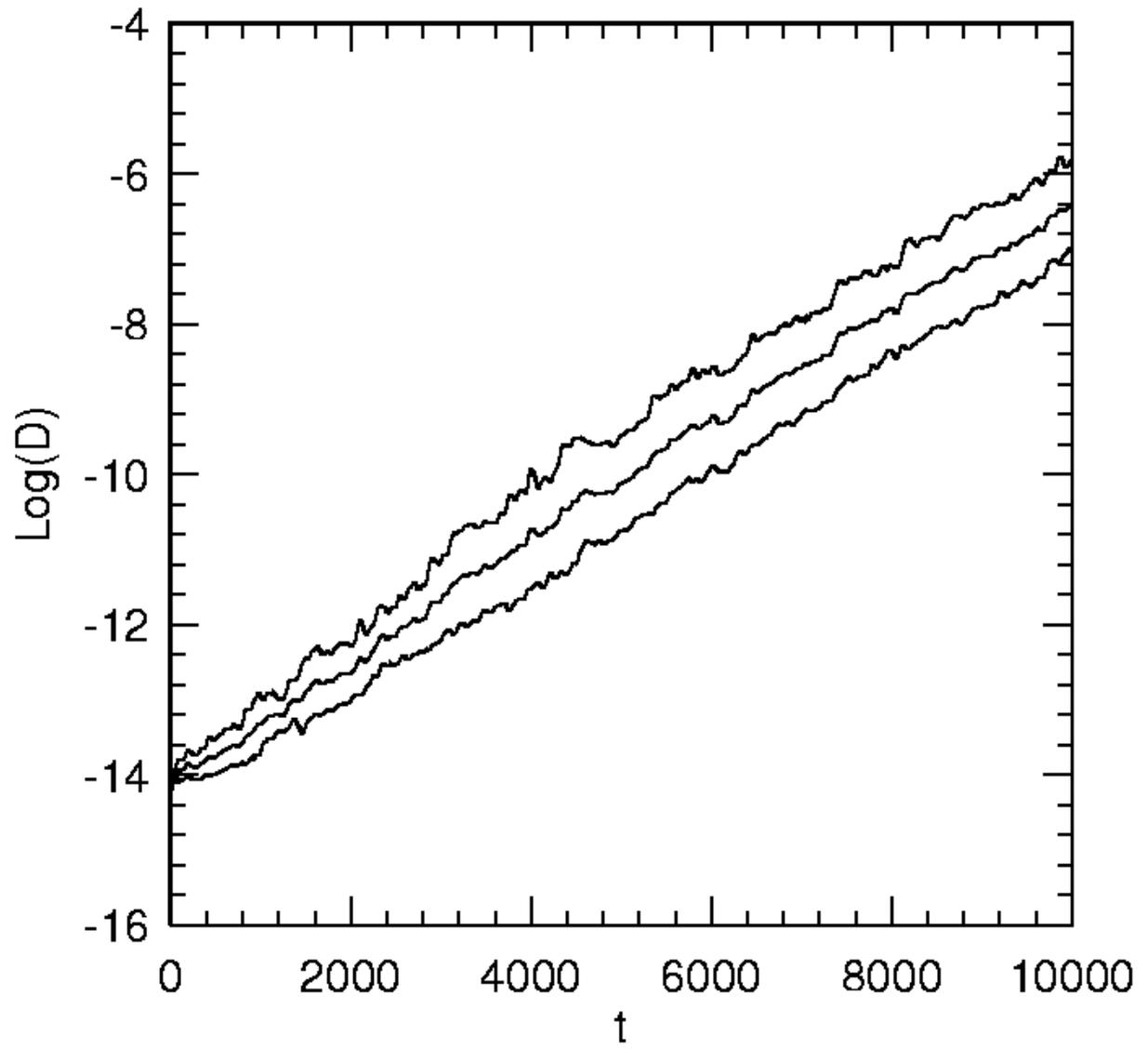
\end{document}